\documentclass[12pt,twoside]{article}   
\usepackage{pdfpages}
\usepackage[super,sort,comma]{natbib}
\usepackage[english]{babel}
\usepackage{fancyhdr}	
\usepackage{authblk}
\usepackage{multirow}
\usepackage{amsmath}
\usepackage{graphicx}
\usepackage{hyperref}
\usepackage[autostyle]{csquotes}
\usepackage{enumerate}
\usepackage{subcaption}
\usepackage{enumitem}
\usepackage[section]{placeins}  
\usepackage[mathlines]{lineno}
\usepackage{setspace}

\makeatletter \renewcommand\@biblabel[1]{$^{#1}$} \makeatother
\newcommand{\cen}[1]{\begin{center} #1 \end{center}}

\newenvironment{packed_item}{
\begin{itemize}
  \setlength{\itemsep}{1pt}
  \setlength{\parskip}{0pt}
  \setlength{\parsep}{0pt}
}{\end{itemize}}

\hypersetup{ colorlinks,
    citecolor=blue,
    filecolor=blue,
    linkcolor=blue,
    urlcolor=blue
}

\usepackage{xcolor}
      
\definecolor{gray}{rgb}{0.6,0.6,0.6}
\definecolor{red}{rgb}{0.85,0,0}
\definecolor{green}{rgb}{0,0.85,0}
\definecolor{blue}{rgb}{0,0,0.85}
\definecolor{beige}{rgb}{0.92,0.87,0.78}
\usepackage[all]{hypcap}    

\begin{document}

\cen{\sf {\Large {\bfseries Quality assurance and reporting for FLASH clinical trials: the experience of the FEATHER trial} \\


\vspace*{8mm}
Isabella Colizzi $^{1,2,*},$
Robert Sch{\"a}fer$^{1,*, \dagger},$
Jonas Br{\"u}ckner$^{3},$
Gaia Dellepiane$^{1},$
Martin Grossmann$^{1},$
Maximilian K{\"o}rner $^{3},$
Antony John Lomax$^{1,2},$
David Meer$^{1},$
Benno Rohrer$^{1},$
Carla Rohrer Bley$^{3},$
Michele Togno$^{1},$
Serena Psoroulas$^{1,4}$
}

$^{*}$Shared first authorship;\\
$^{1}$Paul Scherrer Institute, Switzerland;\\
$^{2}$ETH Zurich, Switzerland;\\
$^{3}$Clinic for Radiation Oncology \& Medical Oncology, University Animal Hospital, Vetsuisse Faculty, University of Zurich, Switzerland \\
$^{4}$University Hospital Zurich, Switzerland\\
}

\pagenumbering{roman}
\setcounter{page}{1}
\pagestyle{plain}
$\dagger$ Author to whom correspondence should be addressed. Email: robert.schaefer@psi.ch

\onehalfspacing
\clearpage
\begin{abstract}
\noindent {\bf Background:} Research on ultra-high dose rate (UHDR) radiation therapy has indicated its potential to spare normal tissue while maintaining equivalent tumor control compared to conventional treatments. First clinical trials are underway. The randomized phase II/III FEATHER clinical trial at the Paul Scherrer Institute in collaboration with the University of Zurich Animal Hospital is one of the first curative domestic animal trials to be attempted, and it is designed to provide a good example for human trials. However, the lack of standardized quality assurance (QA) guidelines for FLASH clinical trials presents a significant challenge in trial design.\\
{\bf Purpose:} This work aims to demonstrate the development and testing of QA and reporting procedures implemented in the FEATHER clinical trial. \\
{\bf Methods:} We have expanded the clinical QA program to include UHDR-specific QA and additional patient-specific QA. Furthermore, we have modified the monitor readout to enable time-resolved measurements, allowing delivery log files to be used for dose and dose rate recalculations. Finally, we developed a reporting strategy encompassing relevant parameters for retrospective studies.\\
{\bf Results:} We evaluated our QA and reporting procedures with simulated treatments. This testing confirmed that our QA procedures effectively ensure the correct and safe delivery of the planned dose. Additionally, we demonstrated that we could reconstruct the delivered dose and dose rate using the delivery log files.\\
{\bf Conclusions:} We developed and used in practice a comprehensive QA and reporting protocol for a FLASH clinical trial at the Paul Scherrer Institute. This work aims to establish guidelines and standardize reporting practices for future advancements in the FLASH-RT field.\\

\end{abstract}

\newpage
\setlength{\baselineskip}{0.7cm}      

\pagenumbering{arabic}
\setcounter{page}{1}
\pagestyle{fancy}
\section{Introduction}
Over the last ten years, significant research has focused on the FLASH effect, which demonstrates that at ultra-high dose rates (UHDR), normal tissue is spared while providing equivalent tumor control compared to conventional (CONV) radiotherapy\cite{favaudon_ultrahigh_2014}. Studies have explored both in-vitro and in-vivo models to understand the underlying mechanisms and to identify the parameters and conditions necessary to achieve this effect. However, the lack of standardization in experimental reporting and procedure makes it challenging to cross-compare experiments and reproduce results \cite{toschini_medical_2025, colizzi_database_2024}. Comprehensive reporting to retrospectively determine optimized FLASH delivery parameters, including dose and dose rate\cite{diffenderfer_current_2022} is critical for successful translation from preclinical studies to clinical applications. 
Accurate reporting is only meaningful when supported by quality assurance (QA) procedures and robust dosimetry that ensure reliable and reproducible doses and dose rate measurements. This is even more important in the context of first clinical trials.

As first trials of UHDR treatments have started, researchers and scientific organizations in the field of radiation oncology have discussed how to define QA goals and requirements for UHDR \cite{taylor_roadmap_2022}. 
A major challenge is that a precise definition of the physics parameters determining the FLASH effect is currently unknown, and consequently, the requirements for QA are also unclear. 
An initial framework for QA and reporting in UHDR clinical trials has been proposed\cite{zou_framework_2023}, highlighting technology gaps and limitations that need to be addressed for the safe implementation of UHDR radiation treatments. Although practical guidelines for machine QA have been developed\cite{spruijt_multi-institutional_2024}, and recommendations outlining minimal and optimal requirements have been published\cite{garibaldi_minimum_2024},    further validation is necessary. To the best of our knowledge, no consensus has been reached yet.

In this work, we aim to present the development and key considerations of QA processes and reporting procedures for the randomized phase II/III FEATHER clinical trial conducted at the Paul Scherrer Institute in collaboration with the University of Zurich Animal Hospital. The procedures were developed in 2023, and the trial was opened for recruitment in 2024. As most of the studies mentioned above were not yet published at the time of development, we had to look for pragmatic but safe solutions to this problem. In reporting our experience, we want to provide the community with a practical example of reporting and QA procedures, which may serve as a model for future clinical and pre-clinical studies.

\clearpage
\section{Methods}
In this section, we will introduce the context of this work, namely the FEATHER trial, which has never been presented before. Further, we will address our framework for QA\cite{zou_framework_2023}, in particular, how we are ensuring that the treatment delivery is (1) safe, (2) as planned, and (3) reproducible. 

\subsection{The FEATHER trial}

The FEATHER trial (FEline orAl squamous cell carcinoma to model human Head\&Neck tumors: A phase II/III randomized trial assessing early toxicity and anti-tumor efficacy of UHDR vs. conventional dose rate proton THERapy) is a curative trial investigating the FLASH effect in feline biopsy-confirmed oral squamous cell carcinoma, jointly run by the University Animal Hospital, Zurich, and the Paul Scherrer Institute. Patients are randomized in a CONV and UHDR arm of proton therapy delivery. To avoid bias related to any assumption on the FLASH effect and its 'threshold dose or dose rate,' only the beam current varies between the two arms; the CONV arm is defined by a proton beam current below 1~nA (0.76 nA at the patient position), and the UHDR arm by the current that maximizes dose rate (382 nA). For reporting purposes, we define the dose rate according to the Folkerts PBS-average dose rate definition \cite{folkerts_framework_2020}. In the conditions above, the dose rate is estimated to be, on average, 0.4~Gy/s for the CONV arm and above 50~Gy/s in the UHDR arm for the expected tumor size ($<$10~cm$^3$). 
Treatment planning and field design are the same in both arms of the study, with no optimization constraints on the dose rate. Each patient receives three fractions of 11~Gy (physical dose) delivered over three consecutive days, amounting to a total of 33~Gy. The treatment utilizes PSI Gantry 1 with transmission proton beams (250~MeV) and pencil beam scanning. Treatment plans incorporate three different beam angles, with each beam optimized to ensure a uniform dose to the target area (SFO), and meet dosimetric requirements and constraints for organs at risk. Each treatment session involves beams from only one angle to minimize the effects of 'split dose' \cite{sorensen_proton_2024}. 

The animals' follow-up takes place at the University Animal Hospital in Zurich, but the arm allocation is unknown to the veterinary radiation oncologists to avoid bias. We examine two key endpoints: the primary is acute toxicity affecting mucous membranes, skin, and other vulnerable organs, such as the eye, and the second is tumor control.

\subsection{Framework for Quality Assurance}
\subsubsection{Safety and dose delivery control}
In clinical trials involving UHDR irradiation, we align safety objectives with those established for standard clinical operations, as recommended by recent literature\cite{garibaldi_minimum_2024, zou_framework_2023, taylor_roadmap_2022}.

For the FEATHER trial, we leverage the clinical infrastructure of PSI Gantry 1 \cite{lin_more_2009}, which successfully treated human patients from 1996 to 2018. We have modified this facility into a UHDR beamline \cite{nesteruk_commissioning_2021}, incorporating essential redundancy and safety measures. However, the high beam currents associated with UHDR delivery introduce unique challenges, particularly the potential for significant dose errors resulting from interlock failures.

The Gantry 1 interlock chain exhibits sufficiently rapid reaction times, allowing us to categorize, following the AAPM TG35\cite{purdy_medical_1993} guidelines, the most probable sources of interlock failure as type-B hazards or lower, following the guidelines of AAPM TG35\cite{purdy_medical_1993}.  As a result, we have not altered the interlock chain. Most of the limits for interlocks are based on dose measurements and, therefore, have not been modified for the trial. We modified only the maximum spot duration check, now defined separately for CONV and UHDR deliveries. This careful approach guarantees safe irradiation delivery for both modalities in compliance with established standards\cite{noauthor_iec_nodate, noauthor_iec_nodate-1} and based on more than 20 years of clinical operation.

It is crucial to note that interlocks and subsequent beam delivery interruptions can influence trial endpoints. Therefore, any interlocks are documented, and if an interlock happens during a UHDR treatment, the patient is excluded from that arm (even though the patient will be followed up as planned).

Regarding dose delivery control in the FEATHER trial, we allow beam currents at the patient location to reach a maximum of 400 nA. We found recombination effects in the dose control monitor chamber (the gantry nozzle ionization chamber) under UHDR up to 20\% at 600 nA\cite{nesteruk_commissioning_2021}. While we must account for these effects, their magnitude does not preclude the use of an ionization chamber for beam current control. Based on our operational experience with Gantry 1 as a UHDR beamline, we observe that charge density and transmission are stable during a full day of operation, although daily variations occur\cite{nesteruk_commissioning_2021}. Therefore, during our QA procedure for UHDR treatment, we calibrate the dose control monitor chamber daily to ensure dose delivery accuracy.

\subsubsection{Quality assurance}
In the FEATHER trial, we optimize patient treatment plans exclusively based on the administered dose, disregarding the dose rate. 
Consequently, from a QA perspective, the primary focus is on ensuring the accuracy of the delivered volumetric dose, which remains invariant across both treatment arms.
We adhere to established QA guidelines for proton therapy using the existing QA procedures from Gantry 1. Since we use a single energy, however, we have reduced the number of QA tests compared to the previous clinical program.

As we expect the patient frequency to be approximately 1/month, we separated the QA testing into:
\begin{itemize}[noitemsep]
    \item Yearly tests;
    \item Treatment-week and patient-specific (PSQA) tests, to be performed within one week of a new treatment as part of the verification procedure of a new starting patient;
    \item Daily tests (DQA) are to be performed only on treatment and verification days.
\end{itemize}

Machine performance and beam model consistency will be evaluated over time as part of the annual QA protocol following standard clinical practices. During the week of patient treatment, we perform checks on the delivery chain, such as verifying the reproducibility of the dose measured by the dose control monitor chamber. \\

\begin{figure}[h!]
    \centering
    \includegraphics[width=0.5\linewidth]{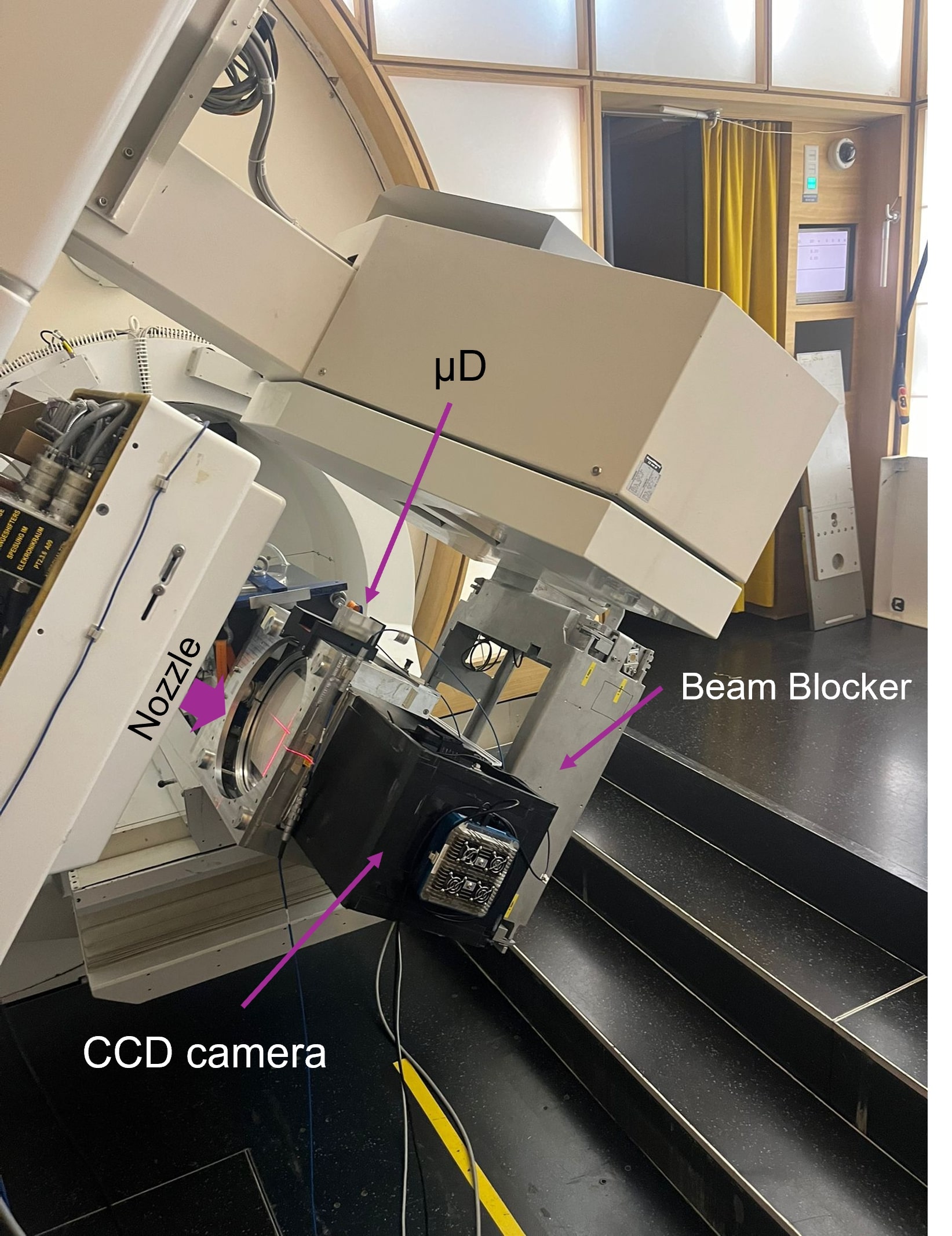}
    \caption{In-house QA phantom designed for the QA procedure: CCD mounted to a rotation stage.}
    \label{fig:phantom}
\end{figure}

\noindent
\textbf{Patient-Specific QA}\\
For patient-specific QA, we designed a measurement device encompassing a CCD camera based on a scintillating foil and a micro diamond detector ($\mu D$)\cite{noauthor_microdiamond_nodate}. All detectors have been verified to be dose rate independent\cite{togno_ultra-high_2022}. The CCD camera and the $\mu D$ are dose cross-calibrated to our institute's reference chamber, traceable to a primary standard. The measurement is performed at a depth of 2 cm, which aligns with the expected depth for most tumors in the trial. The device can rotate so that the measurement plane is always perpendicular to the gantry angle. Figure \ref{fig:phantom} presents a picture of the device. 
The $\mu D$ can be connected to the gantry front-end electronics and measure the current induced by impinging proton radiation. Thanks to the 1~kHz sampled readout of the current, the $\mu D$ can be used for dose rate calculation.

The patient-specific QA consists of the following steps (for each patient field):
\begin{itemize}[noitemsep]
    \item Test of the alignment of the beam isocenter with the cross-hair lasers, which are used for patient positioning. The tolerance is set to 2 mm.  
    \item Evaluation of the spot position and spot size: a 5-spot pattern is delivered, measured with the CCD camera, and compared with the commissioning data. The allowed deviation for positioning is 2~mm, and for beam size (fitted with a 2D-Gaussian) ±10\%. If necessary, a position offset is calculated to correct for any systematic positional shifts. 
    \item Test of the delivered dose, measured with the $\mu D$. The measurement is performed only at the center of the field as the fields are uniform. The measured and planned dose ratio should be within 5\%. If not, the field dose is scaled accordingly. 
    \item Test of the patient field delivery: after eventually applying the corrections defined above, the 2D dose distribution is measured with the CCD. The dose difference between the delivered and the planned dose in the 90\% isodose area is calculated. If the average deviation is larger than 3\%, a boosting factor is determined, and the delivery is repeated. To evaluate differences in the spatial dose distribution, we perform a gamma analysis with distance and dose threshold levels of 3mm/3\%. The pass criterion is met when more than 90\% of the voxels within the 90\% isodose area are at  $\gamma$ $\le$ 1.
    If the agreement is still unsatisfactory, we make a clinical decision based on the location of the largest deviation. 
\end{itemize}

In addition, we measure the local dose rate at different positions within the field using the $\mu D$ to ensure that it is above 40Gy/s in the UHDR case. \\

\noindent
\textbf{Daily QA}\\
The DQA procedure is based on the previously described PSQA. It is oriented, however, toward evaluating machine conditions and safety checks on the day of the treatment.
The dose and especially the dose rate map of the field are very sensitive to phase space and beam position variations; therefore, these parameters are evaluated on a daily basis and compared to the PSQA conditions as recommended by current best practice \cite{taylor_roadmap_2022, garibaldi_minimum_2024}. From experience\cite{nesteruk_commissioning_2021}, we know these parameters are stable within a day of operation. Therefore, we check them only before treatment as part of the DQA procedure. The DQA also includes machine interlock tests to ensure the system responds correctly.

The delivery input file used for daily QA is the one generated during the PSQA procedure. The tolerances used are the same as the ones presented for the PSQA. Additionally, in the case of UHDR delivery, we correct for recombination effects.

\subsubsection{Data recording}
We modified the recording concept used for monitoring in Gantry 1 during patient treatments \cite{lin_more_2009} by establishing a time-resolved data recording system. This setup enables us to generate time traces for reconstructing the delivered dose in a retrospective and time-dependent manner, enabling a post-delivery reconstruction of the dose distribution and dose rate map. The minimum spot length for our UHDR deliveries is 1 ms. Therefore, the data are recorded with a sampling frequency of 1 kHz. The resulting log files contain the following information:

\begin{packed_item}
    \item Delivered dose as a function of time, recorded by two independent dose monitors;
    \item Hall sensor values for two Hall probes positioned in the scanning magnets;
    \item Time stamps at 100~$\mu$s resolution.
\end{packed_item}

Continuous sampling starts as soon as the patient field is loaded for delivery, independently of dose delivery and beam application. This allows for the evaluation of the machine's behavior during beam pauses, including the magnets' ramping and settling time.

We derive the scanning magnet current from Hall sensor readings using a calibrated curve. Subsequently, we translate these magnet currents into lateral spot position coordinates using commissioning data.

We describe each irradiation spot as a two-dimensional Gaussian distribution. The mean positions \(\mu\) are calculated from the Hall sensor values converted into corresponding spot coordinates. The beam widths \(\sigma\) are extracted from commissioning data, while the integral dose is determined from the logged dose values corresponding to dose monitor counts. By summing all individual Gaussian distributions, we generate a 2D dose map of the irradiated field. 

Furthermore, by incorporating the time stamps \(t_{i}\) at the termination of each spot, we can track the temporal evolution of the 2D dose map. We utilize Folkert's metric \cite{folkerts_framework_2020} to then compute the PBS-average dose rate at 5\%-95\% dose levels. This method enables us to calculate the dose rate map for the delivered field across all voxels.
\subsubsection{Reporting}
As the final report of the treatment course, we developed a comprehensive protocol to report information relative to the patient, the machine characteristics, the treatment delivery, the plan evaluation, and relevant information from the QA test, such as the dose, the dose uniformity, and the dose rate. This includes:

\textbf{Patient-specific parameters} \\
Cats with a biopsy-confirmed oral squamous cell carcinoma diagnosis are enrolled under ethical approval and informed consent. The veterinarians collect patient- and tumor-specific characteristics, including the animal's age, breed, sex, histology, and tumor staging information. 

\textbf{Machine-specific characteristics}\\
The beam energy, beam structure (pulse width and frequency), and the currents from both the cyclotron and after the nozzle (calculated knowing the beam transmission) are reported.

\textbf{Setup} \\
The patient positioning, couch, and gantry angles are indicated for each irradiation field.
Anesthetic and patient parameters, including blood oxygen saturation levels of the patient, are recorded before and during irradiation by the veterinarians. During radiotherapy, patients are monitored using a pulse oximeter and capnograph.

\textbf{Dosimetric characteristics}\\
The treatment plan is optimized in our in-house treatment planning system, and the dose is exported as a DICOM file. DVHs, prescription evaluation, and dosimetric information such as DMin, DMean, DMax, V95, D98, and D5-D95 for all the interested organs are reported. Screenshots of the dose distribution are also included. For the PSQA, the dose is recalculated in water, exported as DICOM, and compared with the delivered dose. The difference (from the gamma analysis) and the correction factors (if required) are reported. For comparison, the dose is also recalculated retrospectively with the log files. 

\textbf{Dose rate} \\
In the treatment planning system, we calculate the dose-averaged dose rate\cite{van_de_water_towards_2019} and the PBS-average dose rate\cite{folkerts_framework_2020}. We export the volumetric dose rate distribution as DICOM and the dose rate volume histograms for the most relevant organs field by field. 
Besides the TPS calculation, we calculate the field dose rate by dividing the average dose delivered by the irradiation time, and we measure the PBS-average dose rate at multiple arbitrary points within the irradiation field using the $\mu D$ detector as part of the patient and daily QA and reported in the protocol. Additionally, time information, such as the time between the two irradiation days corresponding to the field-changing time, is included. Moreover, the dose rate can be recalculated per voxel from the log files. This data is not, however, shared with the veterinary radiation oncologists before unblinding. 

\clearpage
\section{Results}

\subsection{Quality Assurance}
\begin{figure}[h!]
    \centering
    \includegraphics[width=\textwidth]{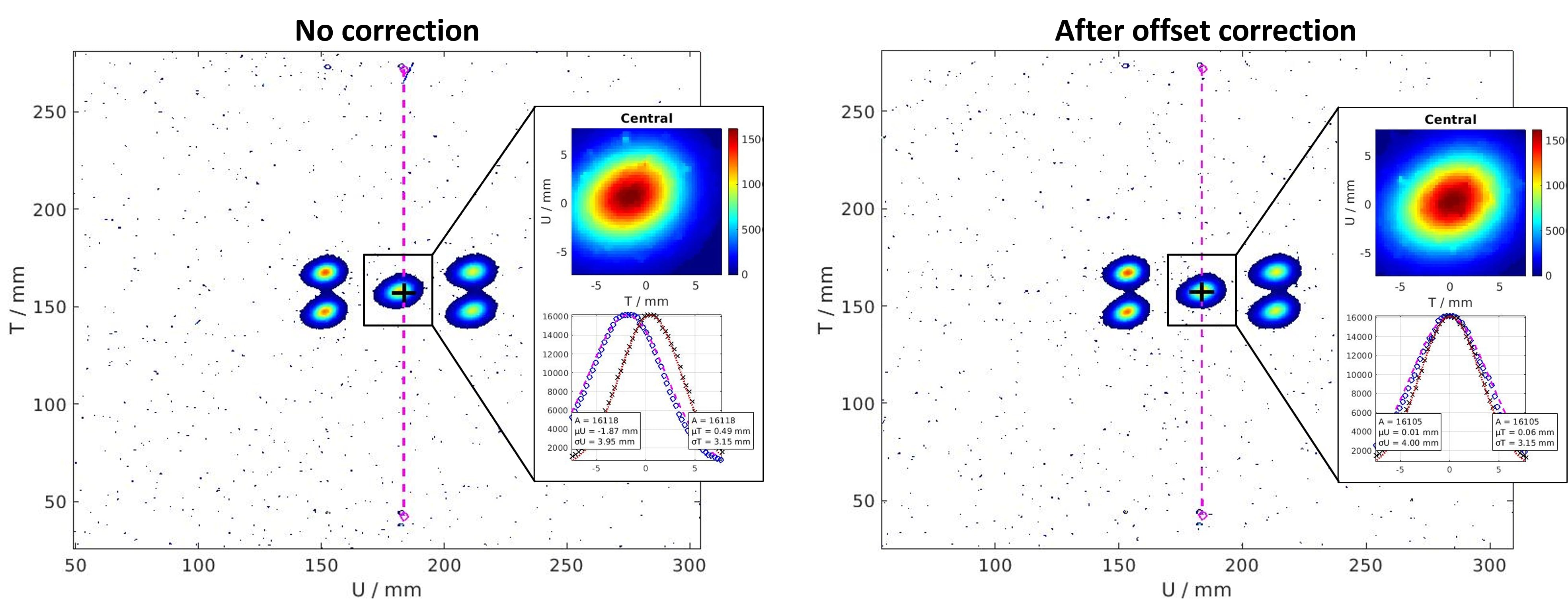}
    \caption{The five-spot pattern is delivered to evaluate spot positioning and beam size (left). The purple line represents the reference central axis that passes through the center of the cross-hair laser. The first delivery is used to evaluate the offset. After calculating the offset, a second delivery with the corrected spot position is performed to confirm the results (right). }
    \label{fig: centering}
\end{figure}

Figure \ref{fig: centering} shows results from the first step of our PSQA and DQA tests: the beam position check. We center the CCD camera mounted on the QA phantom with the cross-hair lasers and deliver the simple five-spot pattern. The central axis is marked on the CCD and is visible during the analysis, serving as a reference for calculating the beam offsets. Once the offsets are determined, we deliver the spot pattern adjusted for these offsets, ensuring that the central spot is now centered with the lasers. Figure \ref{fig: centering} illustrates the five-spot pattern before and after the offset correction.

\begin{figure}[h!]
    \centering
    \includegraphics[width=\textwidth]{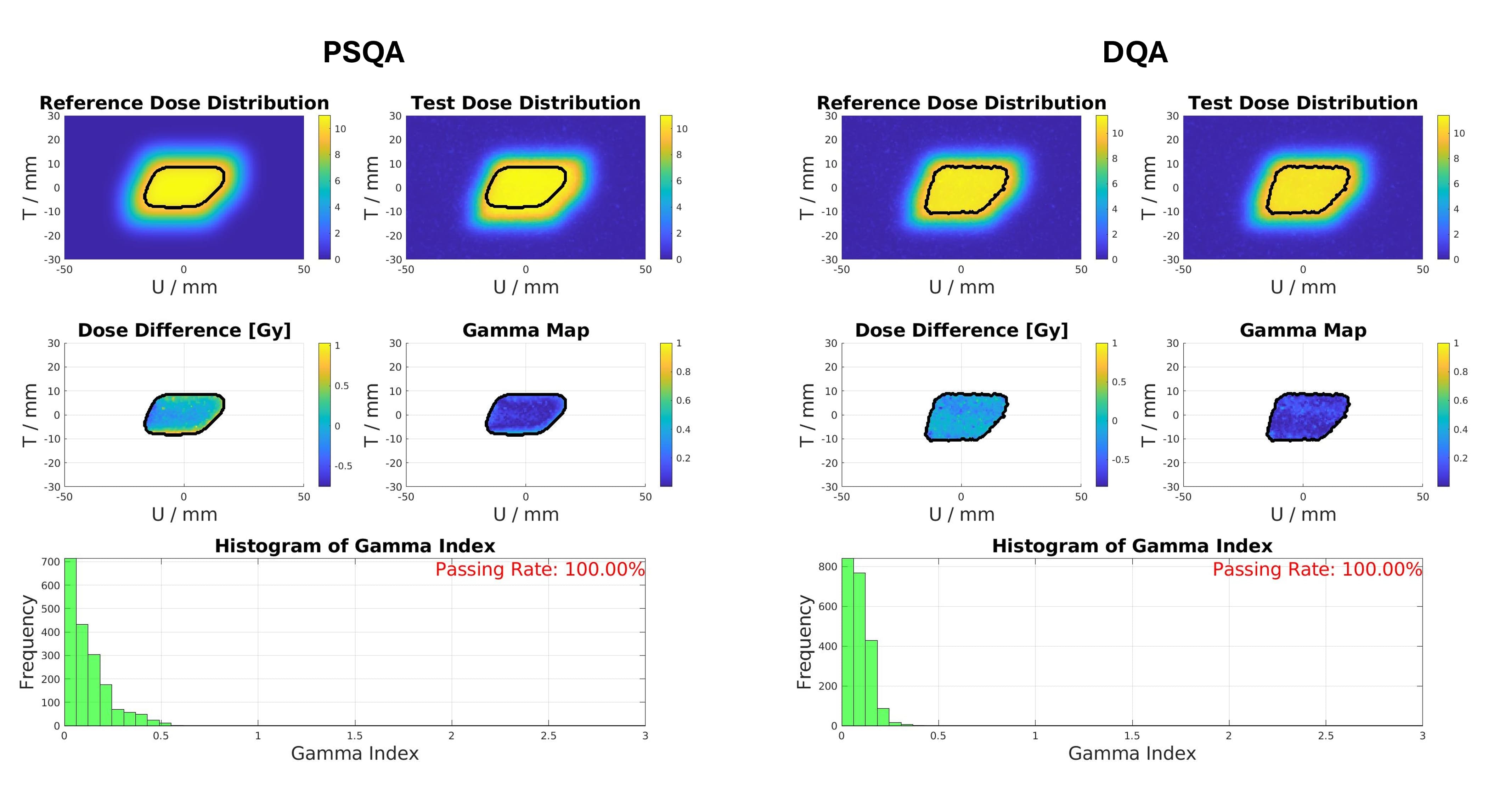}
    \caption{2D dose distribution of the reference and test fields, including the dose difference and gamma analysis between the two, along with an evaluation of the gamma index. On the left, we display the PSQA analysis, where the reference represents the dose distribution recalculated in water from the TPS and the test dose distribution, the measured field. On the right, the reference field is the dose delivered on the day of the PSQA.}
    \label{fig: gamma analysis}
\end{figure}

Figure \ref{fig: gamma analysis} shows the measurements of the 2D dose distributions performed in the PSQA and in the DQA. The  2D dose distribution measured with the CCD camera is compared to the 2D dose distribution recalculated by the TPS in water for the PSQA, or to the measured PSQA dose distribution for the DQA, using average dose difference and gamma analysis for doses above the 90\% isodose level. 

For the presented case in the PSQA (Figure \ref{fig: gamma analysis} (left)), we found a high level of agreement within the 90\% isodose curve. 
This agreement is expected since variations between the beam model imported into our TPS and the actual beam characteristics are more pronounced at the edges of the spot map and at gantry angles close to 0$^\circ$, as the beam transport was optimized \cite{nesteruk_commissioning_2021} for a fixed gantry angle of 90$^\circ$ for the central spot. The TPS cannot accurately simulate these variations, as the beam model remains invariant within the spot map of each scanning angle. 

In the DQA just before treatment, the validated 2D dose distribution from the PSQA is ultimately compared with the daily measurement. Figure \ref{fig: gamma analysis} (right) shows the analysis for this case.
Instabilities in the beam's position or size, even those small enough to pass spot position and size checks, may cause visible deformations.

\begin{figure}[h!]
    \centering
    \includegraphics[width=0.8\textwidth]{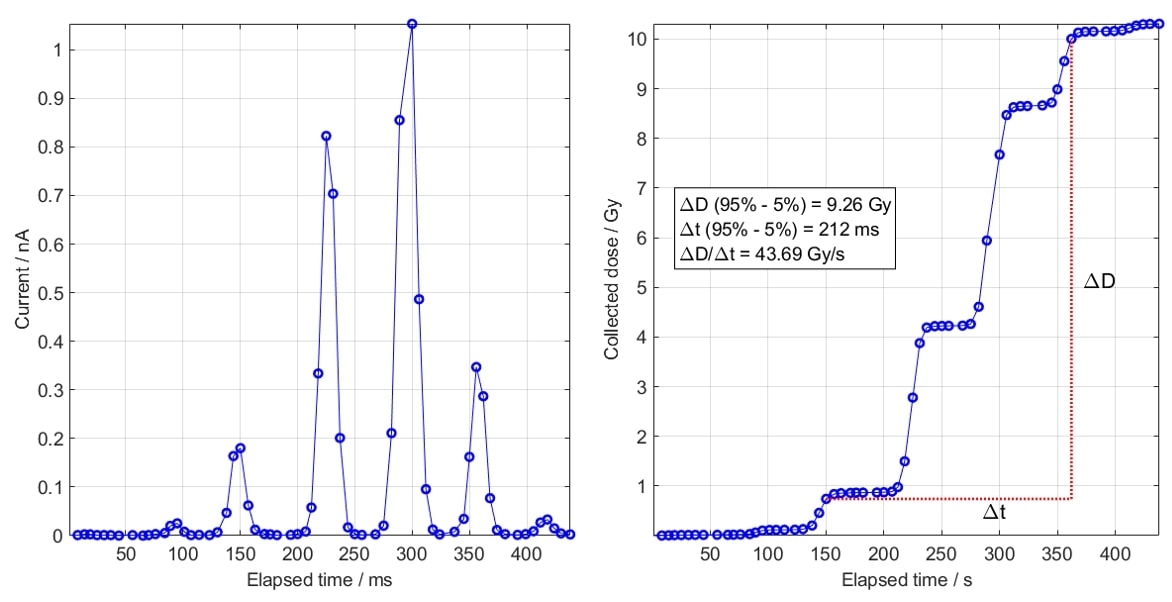}
    \caption{(left) $\mu D$ detector current readout during a UHDR beam delivery. (right) The cumulative sum of the dose delivered.}
    \label{fig: dose rate}
\end{figure}

In addition to dose verification, our workflow includes dose rate measurement at different positions within the field. Figure \ref{fig: dose rate} (left) illustrates an example of $\mu D$ data sampled during a UHDR delivery at a frequency of 1 kHz. To calculate the dose from the measured $\mu D$ output current, we first convert the current to accumulated charge and then to dose using the $\mu$D calibration factor. The time resolution of the readout is sufficient to compute the PBS-average dose rate, as shown in Figure \ref{fig: dose rate} (right).

\subsection{Dose and dose rate reconstruction from log files}
\begin{figure}[h!]
    \centering
    \includegraphics[width=\textwidth]{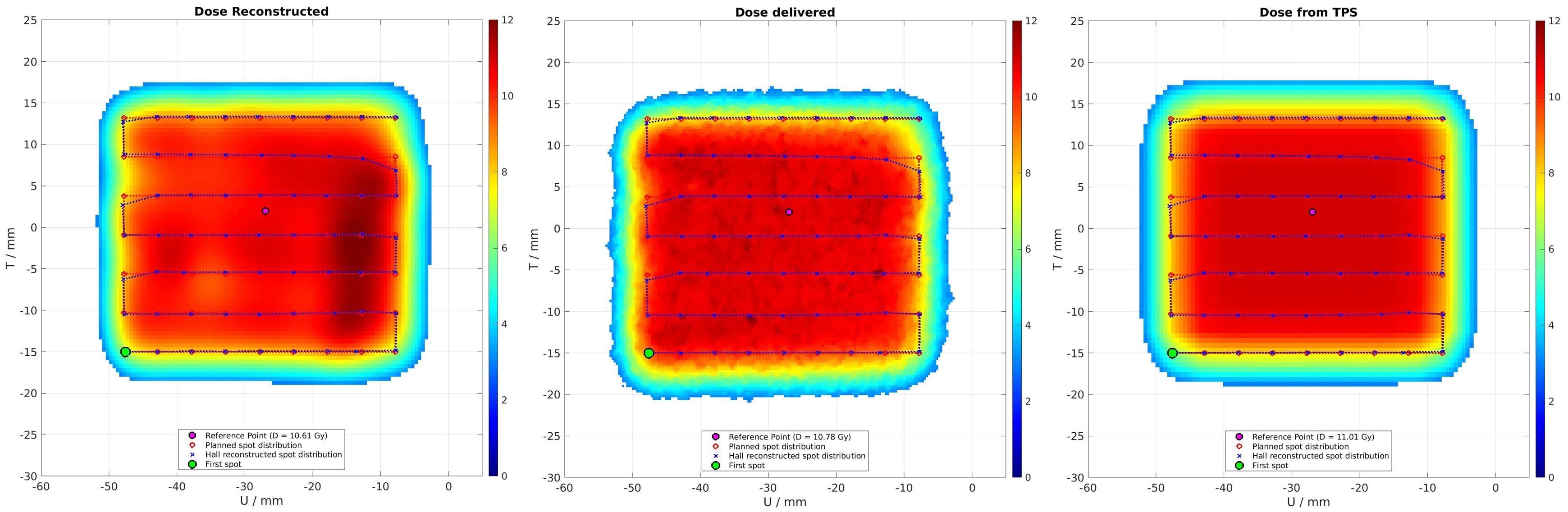}
    \caption{Dose map recalculated from log files (left), measured with the CCD camera (center), and calculated by our TPS (right). The dose measured by the $ \mu $D detector in the reference point is 10.3~Gy. }
    \label{fig: dose map reconstructed}
\end{figure}

In Figure \ref{fig: dose map reconstructed}, we compare, for an example field, the dose reconstructed from the log files with the dose measured with the CCD camera. For reference, we also plot the dose recalculated in water from the TPS. Further, we report the reconstructed, the measured (with CCD), and the simulated (with TPS) dose at an arbitrary position within the treatment field, and we compare it with the dose measured with the $ \mu $D.

It is important to mention that the TPS works with a simplified Gantry 1 beam model. For each gantry angle, we averaged the mean spot size in T and U and the integral (Gy/MU - monitor units), and we assumed these to be constant over the whole map but variable over gantry angles. In addition, we performed dose and dose rate calculations, placing the spots evenly in a rectangular grid. For the dose rate, we assumed a constant spot-changing time in U (4ms) and T (11ms), and constant beam intensity MU/s.   
This simplifies the calculation, as we actually have a variable beam size and integral across the spot map. 
These factors may explain the dose and dose rate distribution variations, particularly at the edges of the spot map.
In addition, the spots in the UHDR mode are not distributed on a rectangular grid, but to reduce the T spot-changing time, the beam is switched on before the planned spot position is reached, as can be observed in Figure \ref{fig: dose map reconstructed} (Hall reconstructed vs. planned spot distribution). 
Moreover, the TPS uses a default cutoff factor for the beam spread to prevent excessive computational demands. 
For the log file recalculation, each spot distribution, represented as a 2D Gaussian, is calculated and then summed across the entire calculation grid.

In Figure \ref{fig: dose map reconstructed}, we observe that the difference between the reconstructed dose and the delivered dose at the reference position is smaller than 2\%, and the measured field shape aligns well with the reconstructed data, indicating good agreement with the reconstructed 2D Gaussian distributions. The calculation from the log files overestimates the dose on the right side of the field. This could be explained by the spot position adjustment in the UHDR delivery required to reduce delivery time. 

\begin{figure}[h!]
    \centering
    \includegraphics[width=0.8\textwidth]{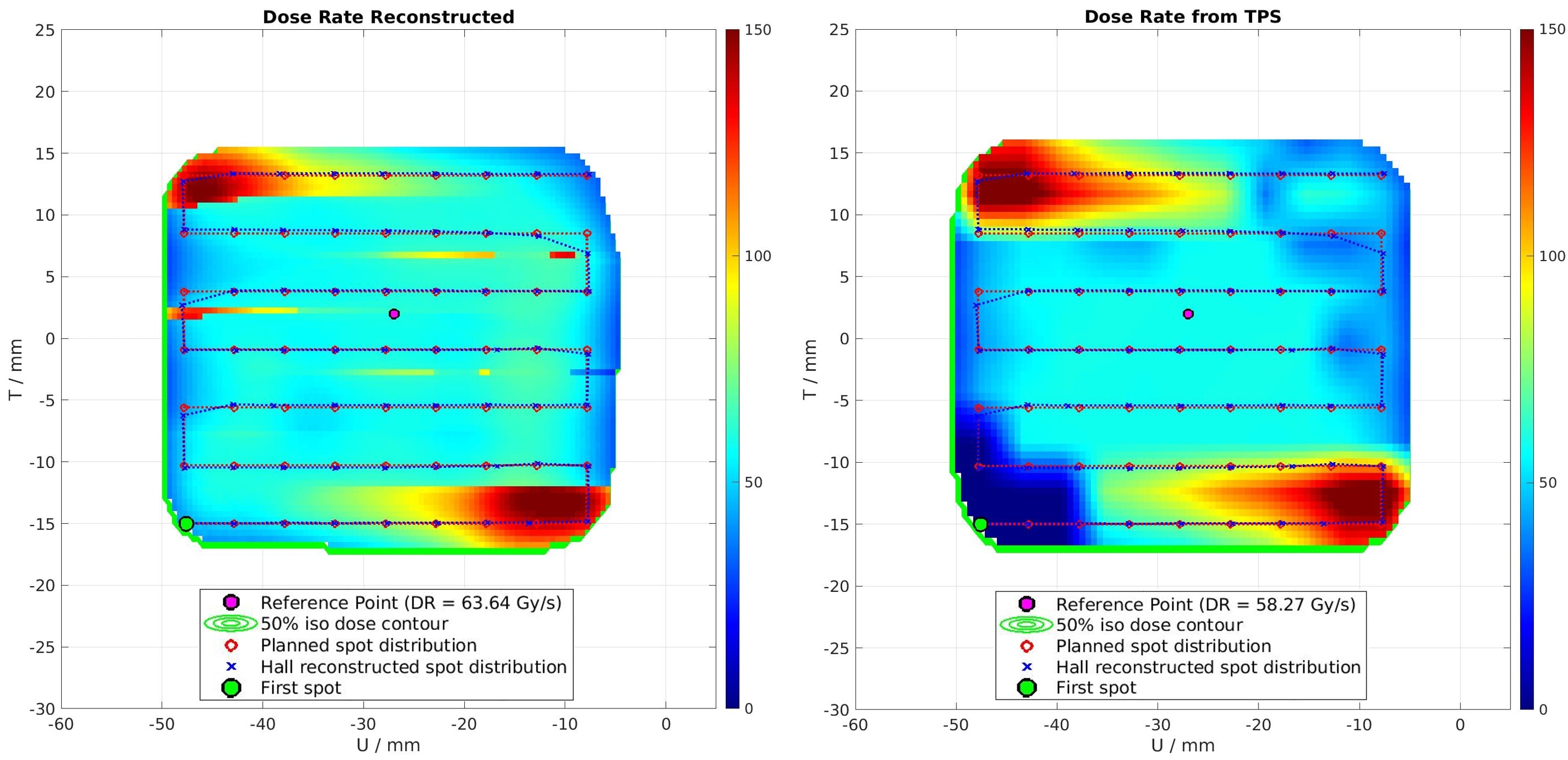}
    \caption{Dose Rate map recalculated from log files (left), and calculated by our TPS (right). The dose rate measured by the $\mu D$ detector in the reference point is 54.3~Gy/s. Additional information are in the Appendix (Figure \ref{fig: FigA1} and \ref{fig: FigA2}) }
    \label{fig: dose rate map reconstructed}
\end{figure}

In Figure \ref{fig: dose rate map reconstructed}, we compare the dose rate reconstructed from the log files with the one simulated by our TPS. At the reference position, the difference in dose rate is within 10\%. 
In the reconstructed dose rate, we observe very sharp variations between neighboring points, particularly at the edges of the field. This is due to the method used to calculate the PBS-average dose rate. The lower or upper threshold (5\% and 95\% of the maximum dose, respectively) in the cumulative dose graph can fall just before or after the plateau, which corresponds to the time when the beam is off and the scanning magnets are changing current for the next spot, as shown in the Appendix (Figure \ref{fig: FigA1}). The position of the starting and endpoint of the calculation impacts the dose rate calculation.  This behavior is not observed in the TPS dose rate calculation, as a step-like function approximates the cumulative dose. In this model, the full spot dose is assumed to be delivered entirely within the beam on time of that spot, with no dose delivered before and after, as illustrated in the Appendix (Figure \ref{fig: FigA1} and \ref{fig: FigA2})).


\subsection{Reporting}
Given the trial's blinded design, the delivery report is divided into three parts, with only the first one made available to the veterinary radiation oncologists before unblinding. The first part includes the dosimetric evaluation of the plan, detailing the 3D dose distribution, prescription evaluation, and dose-volume histograms. The second part consists of dose rate considerations, including measurements with the micro diamond and TPS calculations (dose distribution and dose rate volume histograms). The final part is reserved for the PSQA and DQA.

\begin{figure}[h!]
    \centering
    \includegraphics[width=0.85\textwidth]{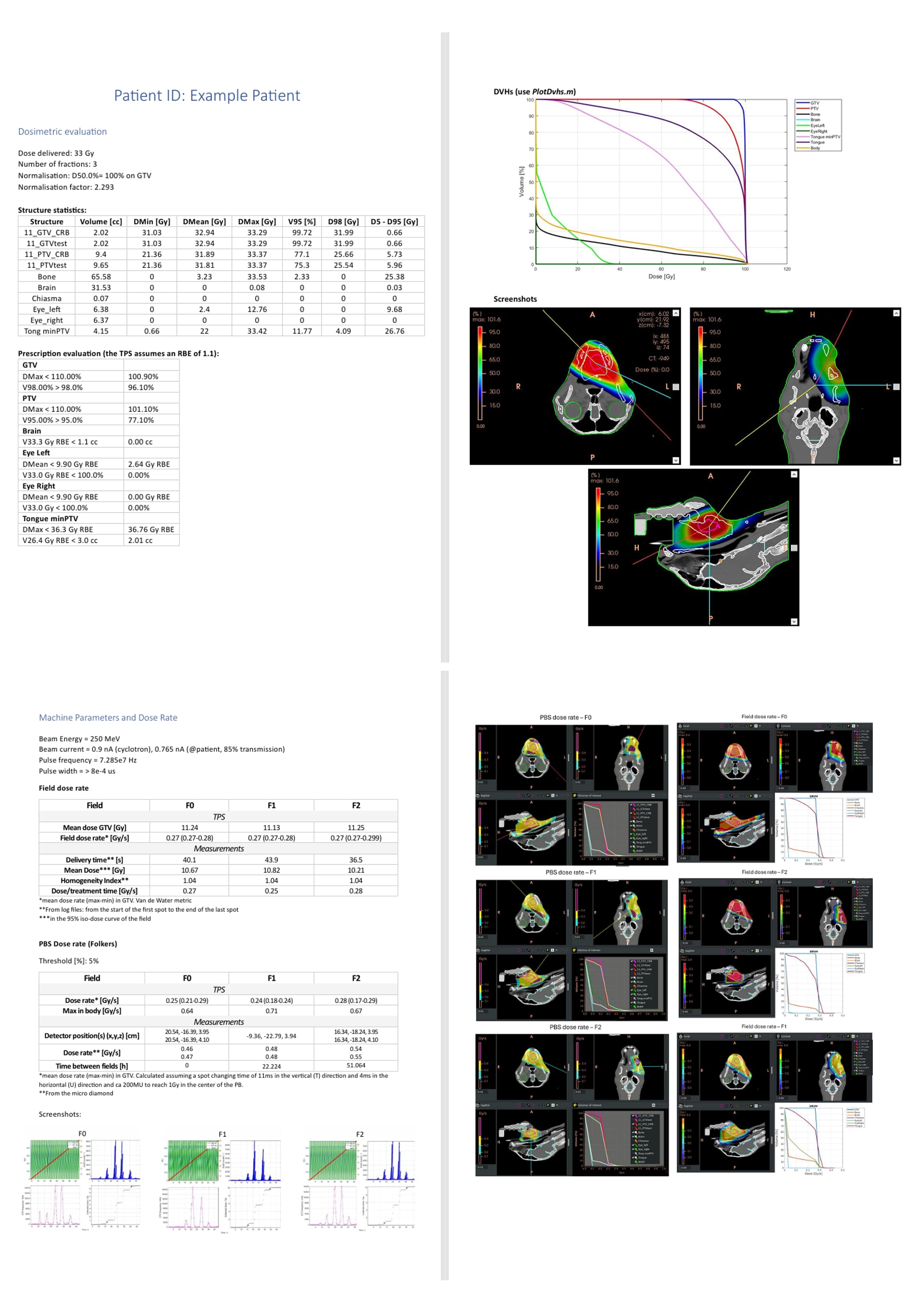}
    \caption{Screenshot of the delivery report for an example patient.}
    \label{fig: report}
\end{figure}

For completeness, we provide the additional materials with a delivery report of a test patient; a preview is shown in Figure \ref{fig: report}.

\section{Discussion}

In this work, we have developed a quality assurance strategy to ensure the safe and accurate delivery of transmission proton treatments in both CONV and UHDR modes, a strategy that has been successfully used during the cat irradiations in the FEATHER trial. Furthermore, we have established a comprehensive protocol encompassing all essential parameters for retrospective analysis and recalculation. To the best of our knowledge, this article provides the first QA and reporting protocol specifically designed for and applied in a FLASH clinical trial.

As part of the QA process of our clinical trial, as we optimize only per dose and not dose rate (the dose rate in the field center should be, in general, above 40Gy/s), our focus was placed on ensuring the accuracy and precision of the dose delivered to the patient. To achieve this, we can refer to the existing QA guidelines for proton therapy. A key advantage of our approach is that it relies on procedures and instruments already used in clinical practice without requiring extensive adaptation or new developments.
We did, however, adapt the readout of our instrumentation to provide a time-resolved dose measurement and allow for log file calculation of time traces. This permits us to recalculate parameters such as dose rate in future analyses if a new definition of the dose rate calculation becomes relevant. With such a complete dataset, we have successfully reproduced the dose and dose rate map with the provided data for a test patient. 
Given the blinded nature of the study, information about dose rate and beam current will be stored in our institutional database and can be disclosed only at the end of the study for each patient. 
No specific and comprehensive guidelines were available regarding what and how to report at the time of our protocol's development. Nonetheless, our protocol closely matches recommendations that were later published  \cite{tobias_bohlen_recording_2024, toschini_medical_2025}.

The main goal of moving preclinical research into clinical trials of UHDR radiotherapy is to evaluate the FLASH effect and assess the feasibility and reproducibility of delivering UHDR treatments. When defining a FLASH trial, however, the first challenge is adapting safety and QA best practices at low to high dose rates. Safety recommendations currently in use\cite{noauthor_iec_nodate}, however, are not adequate for UHDR. In addition, a FLASH trial requires reporting the time behavior of the dose distribution, which in the past was irrelevant in RT.
Work is currently ongoing in international working groups, addressing these open questions \cite{zou_framework_2023,taylor_roadmap_2022,garibaldi_minimum_2024,spruijt_multi-institutional_2024}. With a pragmatic approach, we have demonstrated that many existing clinical procedures can be successfully adapted to meet the requirements of early UHDR clinical trials, where dose rate optimization is either minimal or absent. We believe our approach can apply to studies starting in the coming years since the preclinical knowledge of the FLASH effect does not yet allow its full exploitation. 
Trials in the next years will mainly focus on increasing the beam current and lowering the beam delivery time. In this context, a critical priority should be adapting the monitoring system to allow time-resolved measurements for recording. Given the short irradiation time, attention should be paid to potential errors due to high beam currents. The risk of interlocks occurring during a delivery cannot be entirely eliminated. A better understanding of how beam pauses influence the FLASH effect would be beneficial in understanding how to proceed in the case of interlocks. 

The usefulness of data collected in clinical trials can only be guaranteed by a complete recording and consistent and high-quality reporting. For FLASH deliveries, where the underlying physics and biology parameters are still under intense investigation, completeness of recording is even more important as it guarantees that the data can be further analyzed and interpreted as preclinical knowledge progresses. Time-resolved readouts at very short timescales are crucial to this goal. Dose delivery time traces will be a fundamental part of every future UHDR data analysis and characterization. Both vendors and users of UHDR beams should prioritize this aspect.

\section{Conclusion}
We designed a quality assurance strategy to ensure the safe and accurate delivery of treatments in both CONV and UHDR modes used in the FEATHER trial. Additionally, we developed a comprehensive protocol that includes all necessary parameters for retrospective analysis and recalculation. This approach will guarantee that our data could be used in future comparisons across clinical trials, supporting the transition of FLASH research into clinical application. We hope that our experience can serve as guidance for upcoming trials and that will foster discussions and collaborations within the FLASH community.

\nolinenumbers
\section*{Acknowledgement}
The authors want to thank Martina Bochsler for her support during the irradiation and the Med-IT team, particularly Gabriel Meier, for helping with all TPS-related issues. 

This work was funded by Krebsforschung (Grant number: KFS 5639-08-2022) and partially by the Swiss National Science Foundation (Grant No. 200882)

\nolinenumbers
\section*{Conflict of interest}
The authors have no relevant conflicts of interest to disclose.

\section*{Additional information}
The protocol and procedures employed in the clinical trial were ethically reviewed and approved by the Animal Ethics Council of the Canton of Zurich and Aargau, Switzerland (permit numbers: ZH006/2023). 

\clearpage
\section*{Appendix}
\addcontentsline{toc}{section}{\numberline{}Appendix}
\begin{figure}[h!]
    \centering
    \includegraphics[width=\textwidth]{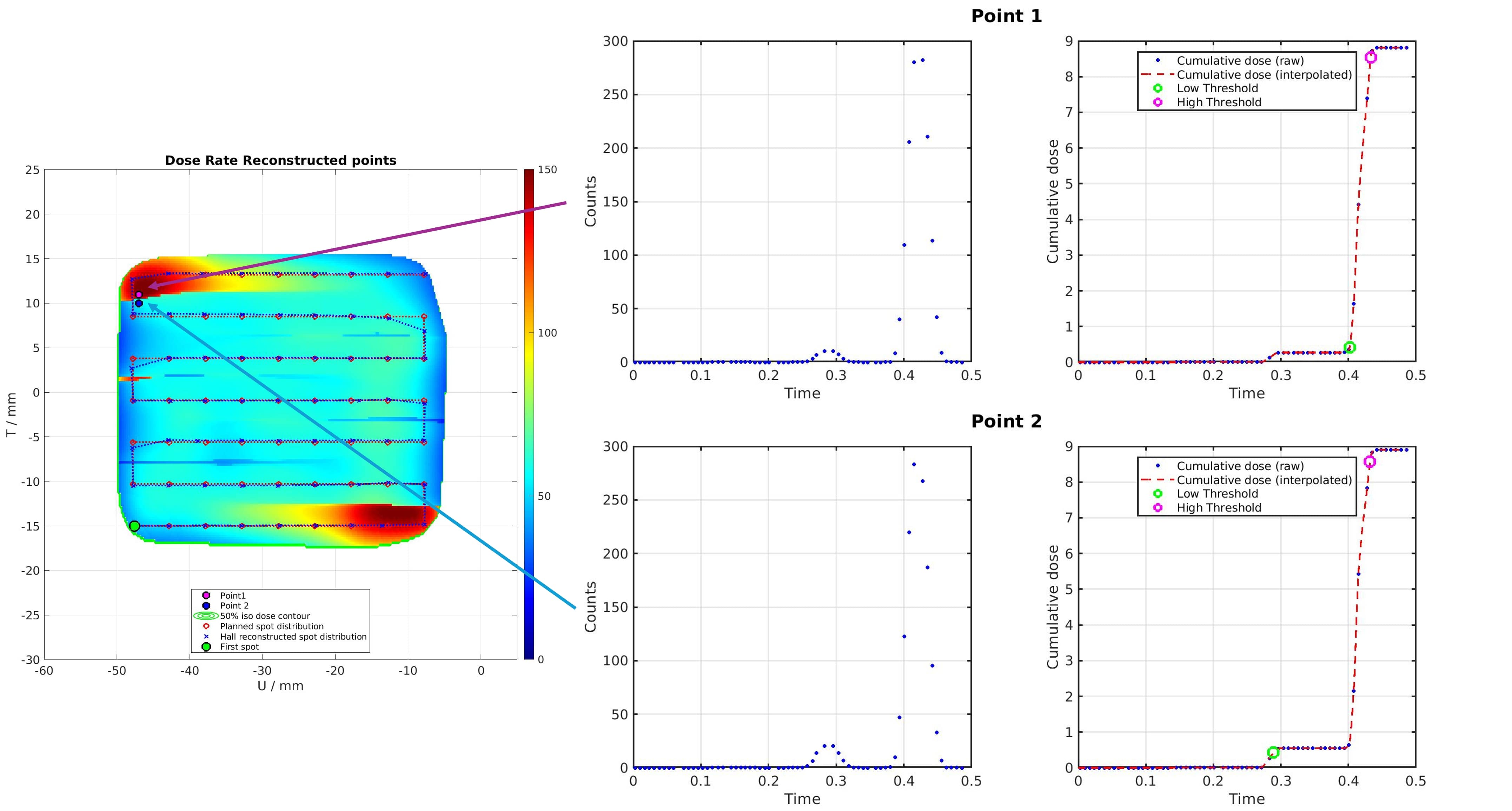}
    \caption{Difference between the dose rate calculation two neighboring points. Even if the count's distribution looks similar as the two points lie close to each other, the position of the lower threshold changes significantly, affecting the dose rate calculation.}
    \label{fig: FigA1}
\end{figure}
\begin{figure}[h!]
    \centering
    \includegraphics[width=\textwidth]{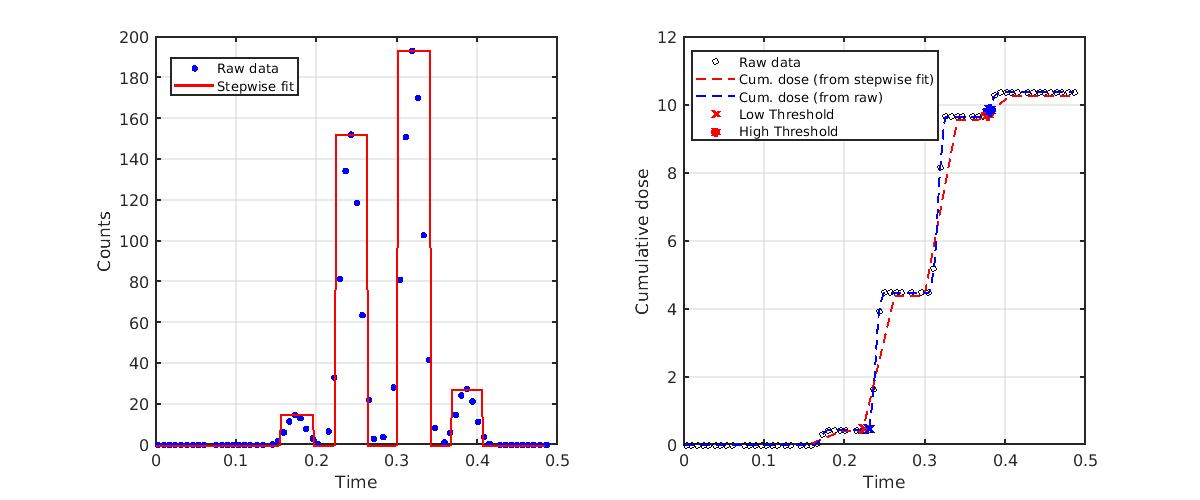}
    \caption{Comparison between log file reconstructed dose rate using a continuous function to represent the counts as a function of time (blue line) and using a stepwise fit of the counts (red line) as an approximation of the TPS calculation.}
    \label{fig: FigA2}
\end{figure}
\clearpage
\includepdf[pages=-]{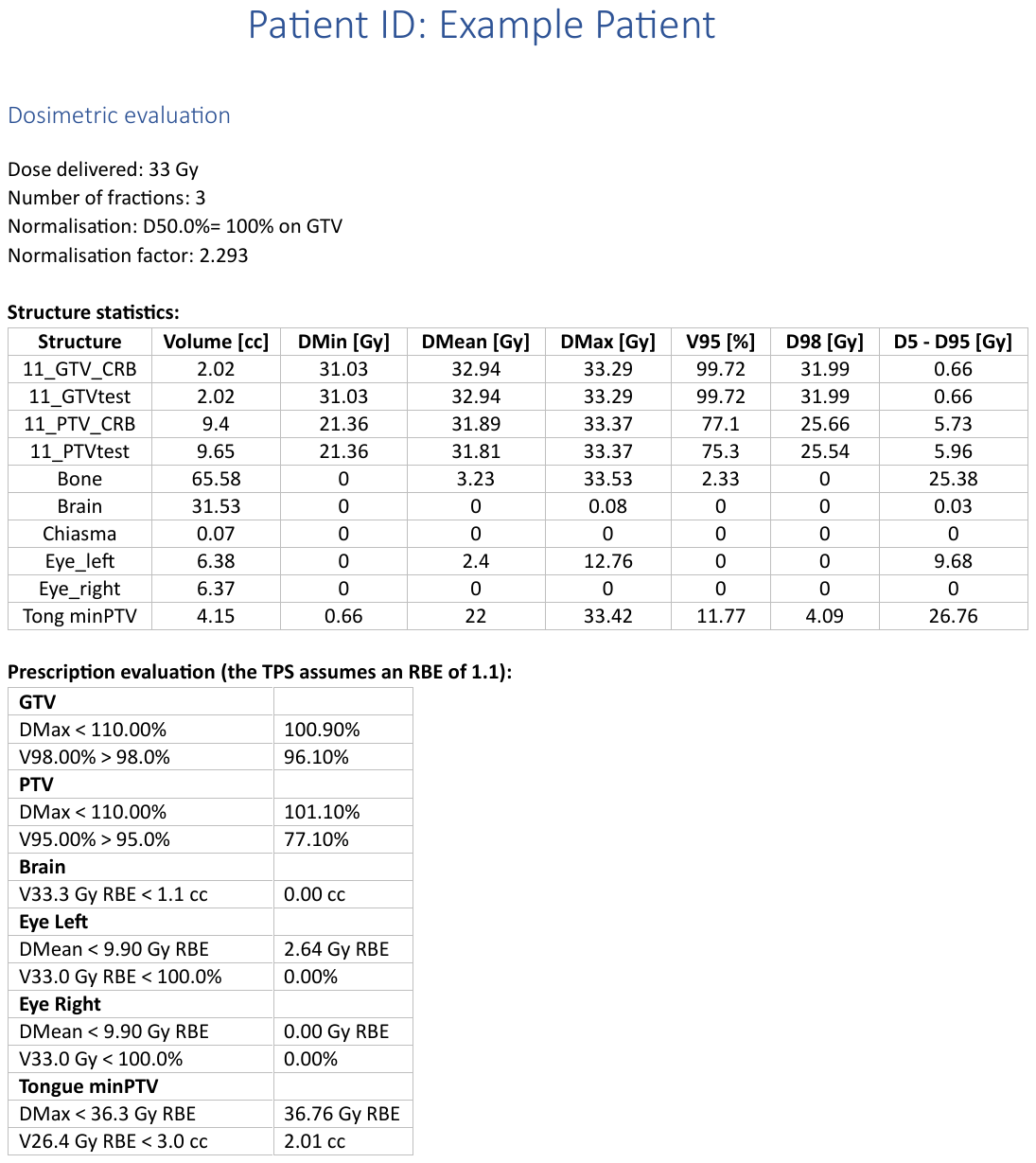}

\clearpage
\section*{}
\addcontentsline{toc}{section}{\numberline{}References}
\vspace*{-20mm}

\bibliography{./references.bib}
\bibliographystyle{medphy}

\clearpage
\listoffigures


\end{document}